\documentclass[prd,twocolumn,showpacs,preprintnumbers]{revtex4}
\usepackage{amssymb,amsmath,mathrsfs,bm}
\begin{document}
\title{The Post--Quasistatic Approximation as a test bed for Numerical Relativity}
\author{W. Barreto}
\email{wbarreto@ula.ve}
\affiliation{Centro de F\'\i sica Fundamental, Facultad de Ciencias,
Universidad de Los Andes, M\'erida, Venezuela.}
\date{\today}
\begin{abstract}
It is shown that observers in the standard ADM 3+1 treatment of matter are the same as the observers used in the matter treatment of Bondi \cite{b64}: they are comoving and local Minkowskian. Bondi's observers are the basis of the post--quasitatic approximation (PQSA) to study a contracting distribution of matter. This correspondence suggests the possibility of using 
the PQSA as a test bed for Numerical Relativity. 
The treatment of matter by the PQSA and its connection with the ADM 3+1
treatment are presented, for its practical use as a calibration tool and as a test bed
for numerical relativistic hydrodynamic codes. 



\end{abstract}
\maketitle

{\sc Introduction}.- 
Unification in the treatment of matter is desirable in Numerical Relativity. 
The standard way to consider matter in ADM 3+1 and characteristic formulations 
leads to flux--conservative equations \cite{nc00}, \cite{sfp02}. These 
procedures are recognized nowadays as Eulerian \cite{font}. Here it is briefly 
reported an unexpected unity provided by an old physical point of view to deal 
with matter \cite{b64}, which combines Lagrangian and Eulerian observers, 
called from now on Bondians. It will be shown below, for the case of
spherical symmetry, that the notion of comoving observers
with the fluid in locally Minkowskian  frames, introduced by Bondi, is exactly the same used in \cite{nc00}. For slow evolution (quasi-staticity) it is possible to extract some general results for an adiabatic fluid starting from reasonable physical assumptions.
Bondi adapted the aforementioned approach to radiation coordinates to study more realistic non--adiabatic problems. 
The Eulerian formulations of Numerical Relativity \cite{nc00}, \cite{sfp02}
actually use Bondian observers in the mathematical treatment of matter.

In 1980 Herrera and collaborators proposed a semi--numeric method \cite{hjr80} 
by elaborating on the original ideas of Bondi, recently interpreted as the post--quasistatic 
approximation (PQSA) \cite{hbds02}. The PQSA starts from any interior static 
solution and leads to a system of ordinary differential equations for quantities 
evaluated at the boundary surface of the fluid distribution. The numerical 
solution of this system allows the modeling of self--gravitating 
spheres whose static limit is the original ``seed'' solution. The approach is 
based on the introduction of conveniently defined effective variables and
heuristic ansatzs, whose rationale and justification become intelligible within 
the context of the PQSA \cite{hbds02}. In the quasistatic approximation, the 
effective variables coincide with the physical variables (pressure and density), 
the method may be regarded as an iterative method with each 
consecutive step corresponding to a stronger departure from equilibrium. 
We show here that the effective variables are exactly the conservative and flux variables
in the standard ADM 3+1 formulation.

Quasi--normal modes are usually employed as a test bed for code calibration  
\cite{qnm}. Hydrodynamic evolution of polytropic spherical
neutron stars can also be used as a test bed for multidimensional
codes \cite{font}. Another possibility is to follow the evolution just
departing equilibrium. The PQSA can be adapted
as a numerical test for more general numerical approaches, not limited
to spherical symmetry \cite{herrera}. Are Bondian observers of relevance for
current codes in ADM 3+1 and characteristic formulations of General
Relativity? The answer could be useful for new practitioners in the area;
we propose the PQSA as a test bed for numerical relativity.

It what follows we briefly review the ADM 3+1 formulation to consider matter. Then, we
demonstrate that Bondian observers are the same as for the ADM 3+1
formulation. Finally, we resume the technical details of the PQSA and propose
how it can be used as a test bed for Numerical Relativity Hydro codes.

{\sc A 3+1 formulation for matter}.- An approach for simulating a self--gravitating, relativistic perfect fluid in spherical symmetry, was documented in \cite{nc00}. Here we present a summary
of that approach to establish its relation with Bondian observers.
Consider a perfect fluid given by the energy--momentum tensor
$T_{ab}=(\rho+p)u_au_b+p g_{ab},$
where $\rho$ is the energy density, $p$ is the pressure, $u^a$ is the 4-velocity and $g_{ab}$ is the spacetime metric. The energy density $\rho$ contains all contributions to the total energy, which for a perfect fluid include the rest mass--energy density, $\rho_0$, and the internal energy density
$\rho=\rho_0 + \rho_0 \epsilon,$
where $\epsilon$ is the specific internal energy. We consider only single--component fluids, and the number density, $n$, is simply related to $\rho_0$ via
$\rho_0=n\mu,$
where $\mu$ is the rest mass of a single fluid particle. The basic equations of motion for the fluid can be derived from local conservation of the
energy--momentum: $T^{ab}_{;a}=0$ and the
particle number: $(nu^a)_{;a}=0$,
where ``;'' is the (covariant) derivative operator compatible with $g_{ab}$. To these conservation laws one must adjoin an equation of state, $p=p(\rho_0,\epsilon)$, which, further, must be consistent with the first law of thermodynamics.

The field equations couple the spacetime geometry, encoded in the Einstein tensor, $G_{ab}$, to the energy--momentum tensor, $T_{ab}$, of the spacetime's matter fields
$G_{ab}=8\pi T_{ab}.$
When using the ADM 3+1 formalism (specialized to spherical symmetry)  to integrate the Einstein equations, and choose the polar--areal coordinates to adopt a polar spherical system $(t,r,\theta,\phi)$, the metric spacetime is written as
\begin{equation}
 ds^2=-A^2dt^2+B^2dr^2+r^2(d\theta^2+\sin^2\theta d\phi^2),
\label{metric}
\end{equation} 
where $A$ and $B$ are functions of $t$ and $r$. In analogy with the usual Schwarzschild form of the static spherically symmetric metric, it is also useful to define the mass aspect function
\begin{equation}
m(t,r)=\frac{r}{2}(1-B^{-2}), \label{ma}
\end{equation}
which coincides with the Misner--Sharp mass \cite{ms64}.

The fluid's coordinate velocity $v$
 \begin{equation}
 v=\frac{Bu^r}{Au^t}
 \label{cv}
 \end{equation}
and the associated Lorentz gamma function, $W=Au^t$,
 are related by
 \begin{equation}
 W^2=\frac{1}{1-v^2}.
 \end{equation}
Defining the conservative variables:
 \begin{equation}
 \tau\equiv(\rho+p)W^2-p=\frac{\rho+v^2 p}{1-v^2},
\end{equation}
\begin{equation}
S\equiv(\rho +p)W^2 v=\frac{(\rho +p)v}{1-v^2}
\end{equation}
and the flux variable
\begin{equation}
\kappa\equiv Sv+p=\frac{p+v^2 \rho}{1-v^2},
\end{equation}
the non--zero components of the energy--momentum tensor are
\begin{equation}
T^t_t=-\tau,\;\;T^r_r=\kappa,\;\;T^t_r=\frac{B}{A}S,\;\;T^\theta_\theta=T^\phi_\phi=p.
\label{Tes}
\end{equation}

For the sake of completeness we write the sufficient set of Einstein equations for the
variables $A$ and $B$ given by the non--trivial component of the momentum constraint (partial differentiation with respect to any coordinate is denoted by a comma) 
\begin{equation}
B_{,t}=-4\pi r A B^2 S \label{mc}
\end{equation}
and by the polar slicing condition, which follows from the demand that metric should have the form (\ref{metric}) for all t:
\begin{equation}
(\ln A)_{,r}= B^2\left[4\pi r\kappa + \frac{m}{r^2}\right]. \label{psc}
\end{equation}
An additional equation for $B$,
\begin{equation}
B_{,r}=B^3\left(4\pi r \tau - \frac{m}{r^2}\right), \label{hc}
\end{equation}
follows from the Hamiltonian constraint.

When the equation of state is not a function of the number density, the time evolution of an ultrarelativistic perfect fluid is completely determined by the conservation of the stress--energy tensor.
The fluid equations of motion can be written in conservative form. We define two
vectors, $\hat{\bf q}$ and $\hat{\bf w}$, which are the conservative and primitive
variables, respectively \cite{nc00},
\begin{equation}
\hat{\bf q}\equiv\left[\begin{array}{c}\tau \\S\end{array}\right],\;\;\;
\hat{\bf w}\equiv\left[\begin{array}{c}p \\v\end{array}\right].
\end{equation}
Also, a flux vector, $\hat{\bf f}$, and a source vector, $\hat{\bf s}$, are
\begin{equation}
\hat{\bf f}\equiv\left[\begin{array}{c}S \\ \kappa \end{array}\right],\;\;\;
\hat{\bf s}\equiv\left[\begin{array}{c}0 \\\sigma\end{array}\right],
\end{equation}
with
\begin{equation}
\sigma=\Theta + 2\frac{A^2 p}{B^2 r},
\end{equation}
where
\begin{equation}
\Theta = A^2 B^2 \left[(Sv-\tau)\left(8\pi r p + \frac{m}{r^2}\right) + p\frac{m}{r^2}\right].
\end{equation}
Clearly in the Minkowskian case $\sigma=2p/r$. We write now the fluid equations of motion in the conservative form
\begin{equation}
\hat{\bf q}_{,t}+\frac{1}{r^2}[r^2 X \hat{\bf f}]_{,r}= \hat{\bf s}, \label{cfe}
\end{equation}
where $X=A/B$.
We will show now how this approach to 3+1 Numerical Relativity is related to Bondi's approach.

{\sc Bondian observers}.-
In order to give physical significance to the variables $\rho$ and $p$,
we follow \cite{b64} to introduce purely local Minkowski coordinates $(T,x,y,z)$ by
\begin{equation}
dT=Adt,\;\;dx=Bdr,\;\;dy=rd\theta,\;\;dz=r\sin\theta d\phi.
\end{equation}
Next we suppose that when viewed by an observer moving relative
to these coordinates with velocity $\omega$ in the radial ($x$) direction,
the physical content of space consists of:
\begin{equation}
\left(
\begin{array}{cccc}
\rho & 0 & 0 & 0 \\
0 & p & 0 & 0 \\
0 & 0 & p & 0 \\
0 & 0 & 0 &  p 
\end{array}
\right).
\end{equation}
A Lorentz transformation readily leads us to
(\ref{Tes}) if $\omega=v$, where
\begin{equation}
\omega=\frac{dx}{dT}=\frac{B}{A}\frac{dr}{dt}. \label{eq_1}
\end{equation}
Now it is clear that $v$ is the local radial velocity and $dr/dt$ is the matter velocity from Bondi's point of view; both are velocities of a fluid element. It makes sense because $u^a=dx^a/ds$ (see eq. (\ref{cv})), so that $dr/dt$ is also a coordinate velocity. Thus, observers that measure $\rho$ and $p$ are comoving with the fluid and local--Minkowskian. Note that we specialize here to the adiabatic case to simplify the presentation and to compare with the ADM 3+1 approach. Also note that $|v|<1$ because this is the velocity 
measured by a Lorentzian observer.

From (\ref{hc}) and (\ref{ma}) we get
\begin{equation}
m_{,r}=4\pi r^2 \tau,
\end{equation} 
which can be easily integrated for any time. Using this integration in
the momentum constraint (\ref{mc}) we obtain 
\begin{equation}
\frac{dm}{dt}=-4\pi r^2 p \frac{dr}{dt},\label{eq_2}
\end{equation}
which is an energy equation (the power), showing clearly how the fluid pressure does work on a material sphere across its moving boundary, as Bondi pointed many years ago. It can be easily shown that this last equation is exactly the first integral of the homogeneous equation of motion in the conservative form (\ref{cfe}). 

The field equation  
$8\pi T^\phi_\phi=8\pi T^\theta_\theta=G^\phi_\phi=G^\theta_\theta$
which reads explicitly
\begin{eqnarray}
&&8\pi p= \frac{1}{B^2}\left[\frac{A_{,rr}}{A}+\frac{A_{,r}}{A}\frac{B_{,r}}{B}+\frac{1}{r}\left(\frac{A_{,r}}{A}-\frac{B_{,r}}{B}\right)\right]\nonumber \\
&&-\frac{1}{A^2}\left[\frac{1}{B}\left(B_{,tt}-\frac{B_{,t}^2}{B}\right)
+\frac{B_{,t}}{B}\left(\frac{B_{,t}}{B}-\frac{A_{,t}}{A}\right)\right],\label{efe}
\end{eqnarray} 
can be written in many ways.
To get some physical insight we can write it as a generalization of the well known
Tolman--Oppenheimer--Volkoff (TOV) equation for hydrostatic support \cite{tov}, \cite{hjr80}, or equivalently, as an equation of motion for the fluid in conservative form (\ref{cfe}), modulo Bianchi identities.
 
We have to satisfy some boundary conditions to match the interior (dynamic) solution with the exterior one, which is static by virtue of the Birkoff theorem. The boundary conditions at some moving radius are that of Darmois--Lichnerowicz \cite{dl}, \cite{hj82}. They are briefly discussed in the next section.

{\sc Post--quasi--static approximation}.-
The PQSA was documented in \cite{hbds02}; it has its origin in the pionering work of Bondi \cite{b64} and in an extension of it known as the HJR method \cite{hjr80}. In this last work the boundary conditions are clearly treated and are crucial to solve the field equations as a system of ordinary differential equations. Here we point out the technical details and its connection with the ADM 3+1 formulation. 

{\it Matching conditions}.- Outside of the fluid distribution, the spacetime is that of Schwarzschild.
In order to match smoothly the two metrics at the surface $r=r_\Sigma(t)$, we require the continuity of the first fundamental form. It follows that
\begin{equation}
A_\Sigma=B^{-1}_\Sigma=(1-2m_\Sigma/r_\Sigma)^{1/2},
\end{equation}
where the subscript
$\Sigma$ indicates the boundary of the distribution and $m_\Sigma$ is the total mass. 
Now, the continuity of the second fundamental form leads us to
\begin{equation}
p_\Sigma=0,\label{pa}
\end{equation}
which expresses the continuity of the pressure at the surface of the matter distribution.
In case of dissipation, that is, heat flow or/and viscosity, the pressure is not continuous anymore \cite{hbds02}.

{\it Identification of effective variables in 3+1}.-
We adapt the PQSA to the contraction of adiabatic spheres,
identifying the conservative and flux variables $\tau$ and $\kappa$
with the effective variables $\tilde\rho$ and $\tilde p$ in \cite{hbds02},
respectively, which are the energy density and pressure in the static limit.
If we know the radial dependence for these variables the hamiltonian 
constraint and the polar slicing condition can be readily integrated to obtain
\begin{equation}
m=\int^r_{0} 4\pi r^2 \tau dr\label{m}
\end{equation}
and
\begin{equation}
\ln \left(\frac{A}{A_\Sigma}\right)=\int^r_{r_{\Sigma}} \frac{(4\pi r^3 \kappa + m)}{r(r-2m)} dr.\label{A}
\end{equation}

Now, we write (\ref{efe}) as
\begin{eqnarray} 
&&\kappa_{,r} + \frac{(\tau + \kappa)(4\pi r^3\kappa + m)}{r(r-2m)}+\frac{2}{r}(\kappa-p)=\nonumber\\
&&\frac{e^{-\nu}}{4\pi r(r-2m)}\left( m_{,tt} +\frac{3m_{,t}^2}{r-2m}-
\frac{m_{,t}\nu_{,t}}{2}\right),  \label{TOV}
\end{eqnarray}
which is the generalized TOV \cite{tov}, \cite{hjr80}.
This equation is exactly the same inhomogeneous
equation of motion for the fluid in a conservative form (\ref{cfe}). To see this
correspondence we have to keep in mind that $\kappa-\tau=p-\rho$ and $\kappa\tau-p\rho=S^2$. We write equation (\ref{TOV}) here to establish a connection between the ADM 3+1 procedure and the PQSA algorithm. It is clear that this TOV is general in the adiabatic context.

{\it Protocol (adiabatic case)}.-
Let us outline here the method \cite{hbds02}:{\it 1}. Take an interior solution to Einstein equations, representing a fluid distribution of matter in equilibrium, with a given $\rho_{st}=\rho(r)$ and $p_{st}=p(r)$. {\it 2}. Assume that the $r$ dependence of $\kappa$ and $\tau$ is the
same as that of $p_{st}$ and $\rho_{st}$, respectively. {\it 3}. Using equations (\ref{m}) and (\ref{A}), with the $r$ dependence of $\kappa$ and $\tau$, one gets $m$ and $A$ up to some functions of $t$, which will be specified below. {\it 4}. For these functions of $t$ one has three ordinary differential equations (hereafter referred to as surface equations), namely, equations (\ref{eq_1}), (\ref{eq_2}) and (\ref{TOV}) evaluated on $r=r_{\Sigma}$. {\it 5}. Once the system of surface equations is determined, it may be integrated for any particular initial data set. {\it 6}. Feeding back the result of integration in the expressions for $m$ and
$A$, these two functions are completely determined. {\it 7}. With the input from the point {\it 6} above, and using the field equations, all the physical variables may be found for any piece of
the matter distribution.

{\it Incompressible fluid as example}.-
Following the PQSA protocol outlined above, we take an interior solution representing a fluid distribution of matter in equilibrium. Now, we recall a Schwarzschild--like model, which corresponds to an incompressible fluid (see \cite{hbds02} for details) in the static case. If a 3+1 matter code is tested  it needs as initial--boundary conditions: $\tau=\tau(t=0,r)$, $S=S(t=0,r)$, $S(t,r=0)=0$. In this example no expansion of $\tau$ near $r=0$
is required because for this model $\tau=3m_\Sigma/4\pi r_\Sigma^3$ for any $r$, in accordance
with incompressibility. Once the system of ordinary differential equations at the
surface is integrated we can, following the protocol, get the ADM 3+1 variable $S$ at the initial slice.
As a matter of fact we can monitor any other geometrical or physical variable
from the PQSA to compare with the ADM 3+1 implementation. This example is not deprived of physical interest because we have enough cummulative evidence that in general the fluid behaves
incompressibly near $r=0$.

{\sc As a test bed}.-
Bondian observers can be at rest at infinity (as in comoving coordinates) or at a local Minkowskian frame. To deal with radiation the proper observers are far from the source ($r\rightarrow\infty$), to deal with matter they are in the bounded source. Bondi's treatment of matter combines Lagrangian local physics with Eulerian global physics in a unambiguous manner. As a final result, it naively looks like purely Eulerian, but here we have shown that it actually is a combination of Lagrangian and Eulerian observers.

We have confidence that the PQSA is a good description up to when the system is just departing from equilibrium. The system always recovers equilibrium by virtue of the PQSA. If stronger departure
from equilibrium --towards collapse-- is desired we have to activate the iterative nature
of the method: the post--post--quasistatic approximation, assuming that the effective variables now share the same radial dependence as that of the physical variables just obtained, and so on. 
In this semi--numerical approach matching across the surface distribution is clearly done, leading to a system of ordinary differential equations which determine the dynamics from initial conditions. The key of the algorithm is an ansatz, based on a specific definition
of the PQSA. Namely, considering different degrees of departure from equilibrium, the post--quasistatic regime (i.e. the next step after the quasistatic situation) is defined as that characterized by metric functions whose radial dependence is the same as that of the quasistatic regime. This in turn implies that some effective variables share the same radial dependence as the corresponding physical variables of the quasistatic regime. Thus, starting with a static
configuration, the first ``level'' off equilibrium, beyond the quasistatic situation, is represented by the post--quasistatic regime. Once the static (``seed'') solution has been selected, the definition of the effective variables together with surface equations allows for
determination of metric functions, which in turn leads to the full description of physical variables as functions of the timelike coordinate, for any region of the sphere. 

The PQSA can be a test bed for a code which must satisfy these conditions: i) smooth matching across the boundary; ii) an equation of state or a relationship between conservative and flux variables (effective variables); iii) an exact departure from the same initial conditions; iv) near the central geodesic, the PQSA is general enough to give the dynamic boundary conditions at any time.

Now we know that in the standard characteristic treatment of matter the physical variables are measured by Bondian observers as well \cite{sfp02}. Because of this new view, we are developing an alternative code which couples matter with scalar radiation, where a PQSA code is used as a tool of calibration and as a test bed. These and other results will be reported elsewhere.
\acknowledgments
Thanks to Beltr\'an Rodr\'\i guez--Mueller, Luis Rosales, Carlos Peralta, Edwin Barrios and Luis Herrera, for their valuable comments.

\end{document}